\documentclass[prl,twocolumn,showpacs]{revtex4}

\usepackage{hyperref}
\usepackage{amsmath,amssymb,tikz}
\usepackage[normalem]{ulem}
\usetikzlibrary{decorations.pathreplacing}

\newcommand{\hatb}{\hat{b}}
\newcommand{\hatal}{\hat{\alpha}}
\newcommand{\hatq}{\hat{q}}
\newcommand{\hatQ}{\hat{Q}}
\renewcommand{\r}{\vec{r}}

\newcommand{\figref}[1]{Fig.~\ref{fig:#1}}

\newcommand{\bra}[1]{\langle #1 |}
\newcommand{\ket}[1]{| #1 \rangle}
\newcommand{\aver}[1]{\langle #1 \rangle}

\begin{document}

\title{Three-point functions in $c\leq 1$ Liouville theory and conformal loop ensembles}

\author{Yacine Ikhlef$^{1,2}$, Jesper Lykke Jacobsen$^{3,4}$ and  Hubert Saleur$^{5,6}$}

\affiliation{${}^1$Sorbonne Universit\'es, UPMC Univ Paris 06, UMR 7589, LPTHE, F-75005, Paris, France}
\affiliation{${}^2$CNRS, UMR 7589, LPTHE, F-75005, Paris, France}
\affiliation{${}^3$Laboratoire de Physique Th\'eorique, \'Ecole Normale Sup\'erieure -- PSL Research University, 24 rue Lhomond, F-75231 Paris Cedex 05, France}
\affiliation{${}^4$Sorbonne Universit\'es, UPMC Universit\'e Paris 6, CNRS UMR 8549, F-75005 Paris, France} 
\affiliation{${}^5$Institut de Physique Th\'eorique, CEA Saclay, 91191 Gif Sur Yvette, France}
\affiliation{${}^6$Department of Physics and Astronomy, University of Southern California, Los Angeles, CA 90089-0484}

\date{\today}

\begin{abstract}

The possibility of extending the  Liouville Conformal Field Theory  from values of the central charge $c\geq 25$ to  $c\leq 1$ has been debated for many years in condensed matter physics as well as in string theory. It was only recently proven that such an extension---involving a real spectrum of critical exponents as well as an analytic continuation of the DOZZ formula for three-point couplings---does give rise to a consistent theory.  We show in this Letter that this theory can be interpreted in terms of microscopic loop models. We introduce in particular a family of  geometrical operators, and, using an efficient algorithm to compute three-point functions from the  lattice, we show that their operator algebra corresponds exactly to that of vertex operators $V_{\hat{\alpha}}$ in $c \leq 1$ Liouville. We interpret geometrically the limit  $\hat{\alpha} \to 0$ of $V_{\hat{\alpha}}$ and explain why it  is not the identity operator (despite having conformal weight $\Delta=0$).

\end{abstract}


\maketitle

Quantum Liouville theory has played a fundamental role in string theory 
and quantum gravity since its introduction by Polyakov \cite{Polyakov}. It also provides a crucial, exactly solvable,  example of {\sl non-rational} conformal field theory (CFT) \cite{Teschner}. These theories are difficult to study, and ripe with unusual features, such as  continuous spectra of critical exponents. They are  however believed to play an important role in many applications, the best known being the transition between plateaux in the integer quantum Hall effect  \cite{Zirnbauer,Roberto}. 

Recent work on Liouville theory has focussed on continuations of the parameters to new domains of values, where new applications may be found, while the extension of the standard CFT  solution presents considerable challenges. We will focus in this Letter on the case of central charges $c\leq 1$, with the Liouville action 
\begin{equation} \label{nuL}
  \mathcal{A} = \int \! \! {\rm d}^2r
  \frac{\sqrt{g}}{4 \pi} \left[\partial_a \phi\partial_b\phi g^{ab}+i\hat{Q}{\cal R} \phi+4\pi \mu{\rm e}^{-2i\hat{b}\phi}\right] \,,
\end{equation}
where $g^{ab}$ is the metric and ${\cal R}$ the Ricci scalar of the underlying 2D space. 
The coupling constant $\hat{b}$ is real, and $\hat{Q}=(\hat{b}^{-1}-\hat{b})$.
The corresponding central charge is $c=1-6\hat{Q}^2$, and the conformal weights of  the vertex operators  $V_{\hatal}\equiv e^{2\hatal\phi}$ are 
\begin{equation} \label{Delta}
  \Delta=\bar\Delta=\hat{\alpha}(\hat{\alpha}-\hat{Q}) \,.
\end{equation}
We note that this CFT is often called ``time-like Liouville'', since, under a redefinition $\phi\equiv i\hat{\phi}$, the dynamics for $\hat{\phi}$ looks like ordinary Liouville except that the kinetic term has negative sign, just like the time coordinate in Minkowski metric. However, since a certain domain of values of $\hat{\alpha}$ is usually implied in time-like theories, we instead refer to (\ref{nuL}) as ``$c\leq 1$ Liouville''.

The extension of Liouville theory to the case $c\leq 1$ was proposed in several contexts. A chief motivation was the hope to compute  correlation functions in non-minimal models of statistical mechanics \cite{Zamolodchikov}  ---for instance, in the percolation problem with $c=0$ \cite{DelfinoViti}---which have remained largely unknown up to this day. An essential step in this extension was the discovery of a formula for the three-point function of vertex operators, generalizing the so-called DOZZ formula for ordinary Liouville \cite{DO,ZZ}, which was the starting point of the construction of the corresponding non-rational CFT. This formula reads \cite{Schomerus,KostovPetkova}
\begin{equation} \label{ZamoC}
  \hat{C}(\hat{\alpha}_1,\hat{\alpha}_2,\hat{\alpha}_3)= \frac{A_{\hatb} \Upsilon(\hatb-\hat{Q}+\hatal_{123}) \widetilde{\prod} \Upsilon(\hatb+\hatal_{ij}^k)} 
  {\sqrt{\prod_{i=1}^3 \Upsilon(\hatb+2\hatal_i)\Upsilon(\hatb-\hat{Q}+2\hatal_i)}} \,,
\end{equation}
where  the product $\widetilde{\prod}$ makes $(ijk)$ run over the three cyclic permutations of $(123)$,
$\hatal_{ij}^k\equiv \hatal_i+\hatal_j-\hatal_k$ and $\hatal_{123} \equiv\hatal_1+\hatal_2+\hatal_3$.
The normalization condition $\hat{C}(\hatal,\hatal,0)=1$ defines $A_{\hat{b}}$. The function $\hat{C}(\hatal_1,\hatal_2,\hatal_3)$ is totally symmetric in its three arguments, and is also invariant under $\hatal_i\to \hat{Q}-\hatal_i$ for any $i$. The function $\Upsilon$ is given by (setting $\hatq\equiv \hatb+\hatb^{-1}$)
\begin{equation} \label{ups}
  \ln \Upsilon(x)\equiv\!\!\int_0^\infty {{\rm d}t\over t}\!\!\left[\left({\hatq\over 2}-x\right)^2\!\! {\rm e}^{-t}-{\sinh^2\big({\hatq\over 2}-x\big){t\over 2}\over \sinh{\hatb t\over 2}\sinh{t\over 2 \hatb}}\right] \,,
\end{equation}
for $0 < \mathrm{Re}(x) < \hatq$. Outside that range, $\Upsilon$ is extended using functional relations; see the Supplementary Material (SM).
$\hat{C}(\hatal_1,\hatal_2,\hatal_3)$ is believed to encode the three-point function of vertex operators, whose general form is dictated by conformal invariance
\begin{equation} \label{VVV3pt}
  \left \langle V_{\hatal_1}(\r_1)V_{\hatal_2}(\r_2)V_{\hatal_3}(\r_3) \right \rangle=\hat{C}(\hatal_1,\hatal_2,\hatal_3) \widetilde{\prod} r_{ij}^{-2\Delta_{ij}^k} \,,
\end{equation}
where $r_{ij} = |\r_{i}-\r_{j}|$.
The two-point function of operators with identical values of $\hatal$ is normalized to have unit residue. 

The consistency of the $c \le 1$ theory with $\hatal={\hat{Q}\over 2}+p$ and $p\in \mathbb{R}$---viz.\ $\Delta=p^2-{\hat{Q}^2\over 4}$ by \eqref{Delta}---was recently demonstrated in \cite{RibaultSanta}, including by extensive numerical checks of crossing symmetry for the corresponding four-point functions. Meanwhile, the relevance of Liouville theory to conformal models of fluctuating loops was pointed out in \cite{Kondev97,Kondev98}. These works, based on the so-called geometrical Coulomb gas (CG) construction \cite{JesperReview}, were however limited to two-point functions. But following the suggestion in \cite{Zamolodchikov}, an interesting proposal was made \cite{DelfinoViti} that the probability for three points to belong to the same Fortuin-Kasteleyn (FK) cluster in the critical $Q$-state Potts model was simply related to the three-point coupling \eqref{ZamoC} for a particular value of the charges $\hatal_i=\hatal(Q)$. 
This was confirmed by numerical simulations \cite{Piccoetc} for real $Q\in [1,4]$, up to a factor $\sqrt{2}$. In \cite{DelfinoViti}, this factor was explained by relating the correlation functions of spin operators to those of kink operators, through Kramers-Wannier duality. Due to the permutation symmetry of the Potts model, the kinks exhibit a two-channel structure in their OPEs, which produces a factor $\sqrt{2}$ in any measurement of the structure constant.

In this Letter we provide a complete  statistical physics interpretation---and extensive numerical checks---of the three-point coupling (\ref{ZamoC}), for continuous values of $c$ and the three independently varying ``electric'' charges $\hatal_i$. Our results apply both to the loop model underlying the $Q$-state Potts model, and to the dense and dilute phases of the O($n$) loop model \cite{Nienhuis82}. In the case of the Potts model, we confirm the factor $\sqrt{2}$ found in~\cite{DelfinoViti}, and we recover a two-channel structure at the level of the transfer matrix Hilbert space.

The context of this physical interpretation is the Conformal Loop Ensemble (CLE) \cite{SheffieldWerner12}, familiar in the study of spin models, CG mappings \cite{Nienhuis,JesperReview}, and more recently the Schramm-Loewner Evolution (SLE) \cite{Sheffield}. The most physical lattice discretization of this ensemble consists \cite{Nienhuis82} in drawing self- and mutually avoiding loops on the hexagonal lattice, with a fugacity per loop equal to a real number $n\in [-2,2]$ and a fugacity per monomer $\beta$ which is $\beta_{\rm c} = (2+(2-n)^{1/2})^{1/2}$ \cite{Nienhuis82} in the {\em dilute} phase, and $\beta_{\rm c} < \beta < \infty$ in the {\em dense} phase; both phases are critical.

To link with the CG approach, we parametrize $n=-2\cos\pi g$, with coupling $g\in (0,2]$, where $g \in (0,1]$ (resp.\ $g \in [1,2]$) describes the dense (resp.\ dilute) phase. Setting $e_0=1-g$ the central charge is $c=1-6 e_0^2/g$ \cite{JesperReview}. This matches $c\leq 1$ Liouville, provided $\hat{b}=\sqrt{g}$, and the link to CLE${}_\kappa$ is $\kappa=4/g$.

The question of which CFT describes the CLE {\em completely} remains open to this day. This CFT should involve a free boson, which one can interpret as the long-wavelength limit of a solid-on-solid model (SOS) dual to the loops. More specifically \cite{BKW76}, the weight $n$ per loop must be understood as arising from oriented loops, which get an extra (complex) weight for every left and right turn. On the hexagonal lattice, this weight is $\exp(\pm i\pi e_0/6)$ so that, after summing over both orientations, and using that the number of left minus the number of  right turns of a closed loop on the hexagonal lattice is $\pm 6$, produces the correct fugacity $n=2\cos\pi e_0$ per loop. The oriented loops define the SOS model by duality, heights on neighboring faces differing by $\pm \Delta h$ depending on the arrow that separates them. Finally, the fact that the weight depends on the number of turns can also be construed \cite{FodaNienhuis,JesperReview} to explain the second, curvature-dependent term in (\ref{nuL}).

The key result of this CG analysis is the understanding of {\em two-point functions}. Specifically, the partition function of the loop model with two special points $\r_1$, $\r_2$---such that loops separating $\r_1$ from $\r_2$ get the fugacity $n_1=2\cos\pi e_1$ (with $e_1$ arbitrary) instead of $n$---decays, for large distances, as $r_{12}^{-2 \Delta(e_1)}$ with
\begin{equation}
  \Delta(e_1) = \frac{e_1^2-e_0^2}{4g} \equiv \hatal_1(\hatal_1 -\hat{Q}) \,.
\end{equation}
In the second equation, we have formally matched the well-known CG result \cite{JesperReview} with the $c\leq 1$ Liouville formula (\ref{Delta}), {\em suggesting} that the (non-local) observable modifying the weight of the loops is related with a vertex operator $V_{\hatal_1}$ of charge
\begin{equation}
  \hatal_1 = \frac{\hatQ}{2}+p, \quad \mbox{with } p= \frac{e_1}{2 \hatb} \,.\label{corresp}
\end{equation}
The symmetry $\hatal_1 \to \hat{Q}-\hatal_1$ then amounts to $e_1\to -e_1$.

The geometrical CG approach is equivalent, in CFT parlance, to a free boson with a charge at infinity \cite{DotFat}. Up to this day, no {\em general} result has been available for the three- and higher-point correlation functions in the loop model. While part of the difficulty lies in determining the correct definition of these correlations in geometrical terms, there are also deep conceptual issues to be overcome. For instance, many non-trivial correlations seem to exist, which the screening construction in \cite{DotFat} would erroneously set to zero for reasons of charge neutrality.

Inspired by the discussion in \cite{Zamolodchikov,KostovPetkova} and the observation in \cite{DelfinoViti,Piccoetc} we have investigated  the possible meaning of the three-point function (\ref{ZamoC}) within the loop model. We have found overwhelming evidence that it admits a geometrical interpretation, which is as follows. 

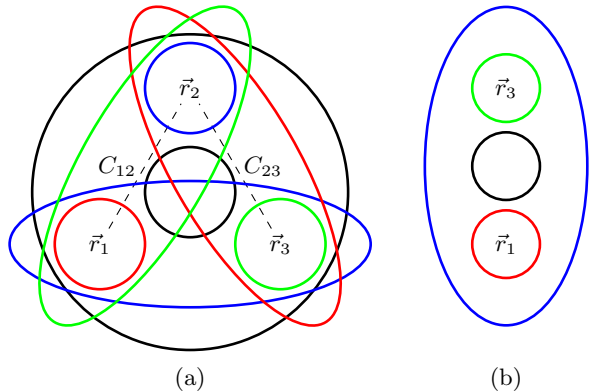
\begin{figure}
  \centering
  \begin{tikzpicture}[scale=0.6]
    \node at (0,0) {$\r_1$};
    \node at (2,3.46) {$\r_2$};
    \node at (4,0) {$\r_3$};
    \draw[black,dashed] (0.2,0.346) -- (1.8,3.114);
    \draw (1,1.73) node[left] {$C_{12}$};
    \draw[black,dashed] (3.8,0.346) -- (2.2,3.114);
    \draw (3,1.73) node[right] {$C_{23}$};
    \draw[red,line width=1pt] (0,0) circle(1.0);
    \draw[blue,line width=1pt] (2,3.46) circle(1.0);
    \draw[green,line width=1pt] (4,0) circle(1.0);
    \draw[black,line width=1pt] (2,1.1533) circle(3.5); 
    \draw[black,line width=1pt] (2,1.1533) circle(1.0); 
    \draw[blue,line width=1pt] (2,0) ellipse (4.0 and 1.4);
    \draw[red,line width=1pt,rotate=120] (0,-3.46) ellipse (4.0 and 1.4);
    \draw[green,line width=1pt,rotate=-120] (-2,0) ellipse (4.0 and 1.4);
    \draw (2,-3) node {(a)};
    \begin{scope}[xshift=9cm]
      \node at (0,0) {$\r_1$};
      \node at (0,3.46) {$\r_3$};
      \draw[red,line width=1pt] (0,0) circle(0.75);
      \draw[green,line width=1pt] (0,3.46) circle(0.75);
      \draw[black,line width=1pt] (0,1.73) circle(0.75);
      \draw[blue,line width=1pt] (0,1.73) ellipse (1.8 and 3.53);
      \draw (0,-3) node {(b)};
    \end{scope}
  \end{tikzpicture}
  \caption{Loop weights in the three-point function. (a) Generic case. Black loops have the bulk weight $n$, while red, blue and green loops have weight
    $n_1$, $n_2$ and $n_3$ respectively. The figure shows all topologies simultaneously; in reality loops cannot intersect. (b) The same, but with point $\r_2$ sent to infinity.}
  \label{fig:loopconfs}
\end{figure}

Consider three points $\r_1,\r_2,\r_3$ in the plane, and imagine we run a cut $C_{12}$ from $\r_1$ to $\r_2$, and another cut $C_{23}$ from $\r_2$ to $\r_3$. We now define a modified  partition function $Z_{n_1,n_2,n_3}(\r_1,\r_2,\r_3)$ of the loop model by giving four different weights to the different topological classes of loops (see \figref{loopconfs}a). If $N_{12} = 0,1$ (resp.\ $N_{23}$) denotes the number of times {\em modulo 2} that a given loop intersects $C_{12}$ (resp.\ $C_{23}$), then its weight is
\begin{equation}
  n_i \,, \quad \mbox{with } i=2 N_{23} + N_{12} \in \{0,1,2,3\} \,. \label{loop weights}
\end{equation}
We have here set $n_0 = n$.
In other words, a loop separating $\r_i$ from the other two points gets weight $n_i$ (with $i=1,2,3$), while a loop surrounding none or all three points gets the bulk loop weight $n_0$. 

Notice that since a loop can be turned around the ``point at infinity'' on the Riemann sphere, we cannot distinguish a
loop surrounding a subset of points from a loop surrounding its complement. Therefore $N$-point functions allow for
$2^{N-1}$ distinct weights. To weigh differently all $2^N$ ways of surrounding subsets of $N$ points, we need to consider an $(N+1)$-point function with $\r_{N+1}$ sent to infinity (in particular no loop can surround $\r_{N+1}$). This is shown in \figref{loopconfs}b for $N=2$.

Parametrizing the weights $n_i=2\cos\pi e_i$ with $e_i \in [-1,1]$, and using (\ref{corresp}) provides a set of three charges $\hatal_i$, with $i=1,2,3$ 
(see SM for higher values of $e_i$). The key idea is that the partition function with weights (\ref{loop weights}) is proportional to the three-point function of the vertex operators $V_{\hatal_i}$ in $c\leq 1$ Liouville. To make this statement definite we need to impose the correct normalization of the partition function. For conciseness we render implicit the insertions $\r_i$ and abbreviate $Z_{n_i,n_j,n_k}(\r_1,\r_2,\r_3) \equiv Z_{ijk}$. We then have our main result that $Z_{123}$ is proportional to (\ref{VVV3pt}), and more precisely
\begin{eqnarray}
  & &
  \hat{C}(\hat{\alpha}_1,\hat{\alpha}_2,\hat{\alpha}_3) = \nonumber \\
  & &
  \quad Z_{123} \sqrt{Z_{000}
    \frac{Z_{011}}{Z_{101}Z_{110}}
    \frac{Z_{202}}{Z_{220}Z_{022}}
    \frac{Z_{330}}{Z_{033}Z_{303}}} \,.
  \label{ratio1}
\end{eqnarray}
The normalization on the left-hand side corresponds to setting $\hat{C}(\hatal,\hatal,0)=1$.

In order to check these formulas, we have devised a method to determine three-point couplings numerically using transfer matrices (see SM for more details).
It turns out more convenient to study models defined on the square lattice, rather than the honeycomb O($n$) model \cite{Nienhuis82} discussed above.
The square-lattice O($n$) model \cite{Nienhuis89} has again dilute and dense phases (representing the same
universality classes). Moreover, we studied the $Q$-state Potts model via its equivalent completely packed O($n=\sqrt{Q}$) model on the square lattice \cite{BKW76},
which produces only the dense universality class.

In our numerical scheme, the axially oriented square lattice is wrapped on a cylinder of circumference $L$.
We split the cylinder into two halves, each consisting of $M \gg L$ rows, and place $\r_1,\r_2,\r_3$ at the bottom, middle and top respectively, all at 
identical horizontal positions. The boundary conditions at the bottom and top are chosen such that neither $\r_1$ nor $\r_3$ can be surrounded by a 
loop. Then (\ref{loop weights}) amounts to giving weight $n_1$ (resp.\ $n_3$) to non-contractible loops below (resp.\ above) $\r_2$,
and weight $n_2$ to contractible loops that surround $\r_2$. All other loops get weight $n$.
We then determine numerically the corresponding partition functions by acting with the transfer matrix, and form the ratio corresponding
to (\ref{ratio1}) after the conformal map from the plane to the cylinder:
\begin{equation}
  \hat{C}(\hat{\alpha}_1,\hat{\alpha}_2,\hat{\alpha}_3) = \frac{Z^{\mathrm{cyl}}_{123}}{Z^{\mathrm{cyl}}_{220}}
  \sqrt{\frac{Z^{\mathrm{cyl}}_{202} Z^{\mathrm{cyl}}_{000}}{Z^{\mathrm{cyl}}_{101} Z^{\mathrm{cyl}}_{303}}} \,, \label{ratiocyl}
\end{equation}
where $Z_{ijk}^{\mathrm{cyl}}$ is the analog of $Z_{ijk}$ on the cylinder, with the points $\r_1$ and $\r_3$ sent to the two extremities of the cylinder.
We finally compare the values of this ratio with the Liouville result (\ref{ZamoC}), evaluated as described in SM. We have found excellent agreement for a wide range of values of the four loop weights. We shall now illustrate this by reporting in detail on two cases which are particularly noteworthy.

\begin{figure}
  \vskip-0.6cm
  \includegraphics[scale=.34]{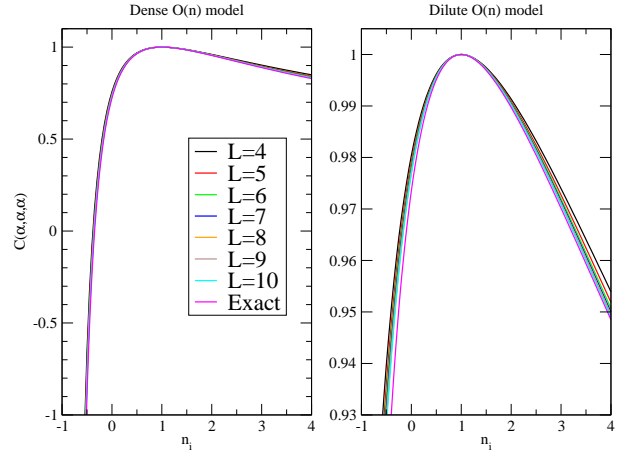}
  \vskip-0.9cm
  \caption{$\hat{C}(\hatal,\hatal,\hatal)$ as a function of $n_1=n_2=n_3$ in the dense and dilute O($n$) model with $n=1$.
  }
  \label{fig:Caaa}
\end{figure}

The \underline{first case} corresponds to $n_1=n_2=n_3$, for which \figref{Caaa} shows results 
both for the dense ($c=0$, percolation) and dilute ($c=\frac12$) phases of the O($n$) model
with $n=1$. The agreement with (\ref{ZamoC}) is excellent for the whole range of $n_i$, and can be further
improved by finite-size scaling (FSS) extrapolation (see SM).
The divergence $\hat{C} \to -\infty$ as $n_i \to -1^+$, due to a {\em pole} in the theoretical formula,
is beautifully reproduced by the numerics. Note that to reach values near $n_i = -1$ requires analytical continuation of $\Upsilon$ beyond the domain
of validity of the integral representation (\ref{ups}); see SM for details. The formula (\ref{ZamoC}) works perfectly well for $n_i>2$ as well,
corresponding to {\em imaginary values} of the charges $\hatal$.
We conjecture that (\ref{ZamoC}) applies indeed for all $\hatal_i \in \mathbb{C}$, corresponding in general to complex values of $n_i$.
For $n \in \mathbb{R}$, we need $-2\leq n \le 2$ to ensure a critical theory; nothing is known about  complex values of $n$.

We stress that for generic $(n_1,n_2,n_3)$, the partition function $Z_{n_1,n_2,n_3}$ cannot be encoded in the local vertex model equivalent \cite{JesperReview} to the O($n$) model. Indeed, in the former, $Z_{n_1,n_2,n_3}$ can only be obtained by introducing a twist factor ${\rm e}^{\pm i\pi e_1}$ (resp. ${\rm e}^{\pm i\pi e_3}$) associated to the arrow flux through the cut $C_{12}$ (resp. $C_{23}$), and this forces $n_2$ to take one of the four values $-2\cos\pi (g \pm e_1 \pm e_3)$. In the CG approach \cite{DotFat}, this corresponds to a three-point function satisfying the neutrality condition with one type of screening charges. Hence, going beyond this case is only possible in the non-local loop model.

Repeating the numerics for the loop model associated to the Potts model, we still find excellent agreement for $n_i > 0$. For $n_i < 0$ we have very large FSS effects, but going to large sizes ($L=16$) the extrapolation still gives a decent agreement with the theoretical formula. Right at $n_i = 0$, and only there, one can check with great accuracy that the numerical data converge to $\sqrt{2}\hat{C}$ (and not to $\hat{C}$ as usual). This is precisely the case \cite{DelfinoViti,Piccoetc}, where (\ref{VVV3pt}) is interpreted as the probability that $\r_1,\r_2,\r_3$ belong to the same FK cluster. Actually, one can show \cite{J14,GRS} that the space of states on which the transfer matrix is acting splits, when $n_i=0$ and {\sl only} then, into two isomorphic subspaces, which are technically irreducible representations of the lattice (periodic Temperley-Lieb) algebra underlying the dynamics of the model; see SM for more details. Numerical methods and their associated normalizations measure the three-point constant within one subspace only (selected by the boundary conditions imposed at the ends of the cylinder), while, by analyticity, the Liouville result holds for the whole space of states. This leads, by easy considerations on (\ref{ratiocyl}), to a measured three-point constant which is too large by a factor $\sqrt{2}$ for $n_i = 0$.

The \underline{second case} we consider in some detail here is when one of the charges vanishes, say $\hatal_2 = 0$. The vertex operator then has $\Delta_{\hatal_2} = 0$. In ordinary CFT this would imply $V_{\hatal_2}(\r_2) = I$, the identity operator, so that (\ref{VVV3pt}) reduces to a two-point function, equal to zero by conformal invariance, unless $\hatal_1=\hatal_3$. However, (\ref{ZamoC}) does not exhibit this feature at all, and the function $\hat{C}$ remains highly non-trivial even when one of the arguments vanishes \cite{McElgin}. We still have $\hat{C}(\hatal,0,\hatal)=1$, compatible with the normalization of two-point functions, but in general $\hat{C}(\hatal_1,0,\hatal_3) \neq 0$. This will be illustrated numerically below. 

To discuss the corresponding geometrical meaning, we consider the loop model with $n_2=n$. According to \figref{loopconfs}a the weight of a loop encircling $\r_1$ (or $\r_3$)
depends on whether it {\em also} encircles $\r_2$. Thus, $\r_2$ is not invisible at all, and $V_{\hatal_2}(\r_2)$ is in fact a {\em marking}
operator, distinct from $I$, even though $\Delta_{\hatal_2} = 0$. The three-point function (\ref{VVV3pt}) is then
\begin{equation}
  \frac{Z_{103}}{Z} \approx \hat{C}(\hatal_1,0,\hatal_3) \left( \frac{r_{12}}{r_{23}} \right)^{\Delta_{\hatal_3}-\Delta_{\hatal_1}}
  \left( \frac{a}{r_{13}} \right)^{\Delta_{\hatal_1}+\Delta_{\hatal_3}} \nonumber
\end{equation}
where we have, for the time being, considered an unnormalized quantity, where the (unknown) lattice cutoff $a$ appears explicitly.
This depends on $\r_2$ (despite $\Delta_{\hatal_2} = 0$), implying a non-zero derivative wrt $\r_2$. Therefore $L_{-1}$ acts non-trivially
on $V_{\hatal_2}$, and this field is {\em non-degenerate}. We now send $\r_2\to\infty$.
The fraction of loops encircling both $\r_2$ and at least one of the other two points becomes negligible, so we get rid of the marked point. The first scale factor above disappears, and we can then get rid of the second one by normalizing with the two-point function. Hence
\begin{equation}
  {Z^{\mathrm{cyl}}_{n_1,n_3}\over\sqrt{Z^{\mathrm{cyl}}_{n_1,n_1}Z^{\mathrm{cyl}}_{n_3,n_3}}} = \hat{C}(\hatal_1,0,\hatal_3) \,, \label{Ca1a3}
\end{equation}
where $Z^{\mathrm{cyl}}_{n_i,n_j}$ now denotes the partition function where loops encircling $\r_1$ (resp.\ $\r_3$) get weight $n_1$ (resp.\ $n_3$), while those encircling both points get weight $n$. 

Alternatively, we can keep $n_2$ generic (so that $\Delta_{\hatal_2} \neq 0$), send $\r_2 \to \infty$ as before, but choose the microscopic boundary conditions so that no loop can encircle $\r_2$. Loops encircling both $\r_1$ and $\r_3$ then get the non-trivial weight $n_2$, and the ratio (\ref{Ca1a3}) produces $\hat{C}(\hatal_1,\hatal_2,\hatal_3)$. This situation is depicted in \figref{loopconfs}b.

We can also simplify (\ref{Ca1a3}) even further by noticing that $\hat{C}$ remains non-trivial even when {\em two} of the charges $\hatal_i$ vanish. We have then
\begin{equation}
  \frac{Z_{n_1,n}^{\mathrm{cyl}}}{\sqrt{Z^{\mathrm{cyl}}_{n_1,n_1}Z^{\mathrm{cyl}}}}= \hat{C}(\hatal_1,0,0) \,.
\end{equation}
Here, $Z^{\mathrm{cyl}}_{n_1,n}$ is the partition function where loops encircling $\r_1$ but not $\r_3$ get a weight $n_1$.  

\begin{figure}
  \vskip-0.6cm
  \includegraphics[scale=.34]{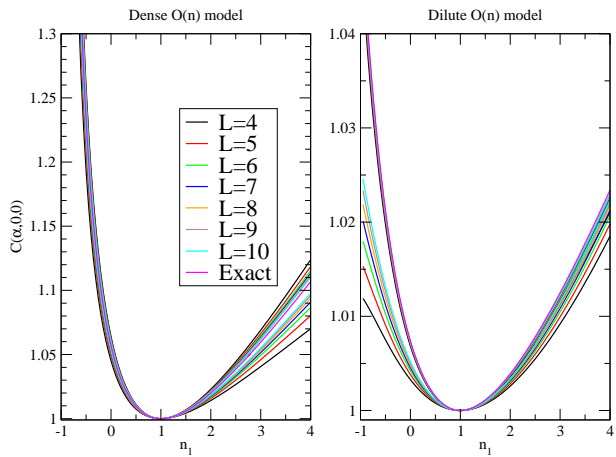}
  \vskip-0.9cm
  \caption{$\hat{C}(\hatal_1,0,0)$ and $\hat{C}(0,\hatal_1,0)$ as functions of $n_1$ in the dense and dilute O($n$) model with $n=1$.
    The two positions of the vertex operator $V_{\hatal_1}$---at one extremity or in the middle of the cylinder---give different microscopic
    results, but their $L \to \infty$ limits agree with the same analytical formula.
  }
  \label{fig:Ca00}
\end{figure}

We have checked these interpretations of $\hat{C(}\hatal_1,0,\hatal_3)$ and $\hat{C}(\hatal_1,0,0)$ numerically. The latter case is shown in \figref{Ca00}.
For the Potts model one obtains similar results, except that now $\hat{C}/\sqrt{2}$ is observed when $n_1 = 0$, as can again be seen from (\ref{ratiocyl}) and the splitting of the state space.

Summarizing, the loop model with four different weights (see \figref{loopconfs}a) gives a consistent interpretation of the three-point function in $c\leq 1$ Liouville. It is also consistent with interpreting $\hat{C}$ as an {\em OPE coefficient}. This can be seen geometrically by considering the limit $r_{23}\to 0$. The amount of loops with weight $n_2$ or $n_3$ is then negligible, since they would have to be ``pinched'' between $\r_2$ and $\r_3$. Instead, loops get weight $n_1$ whenever they encircle $\r_1$ or $\r_2$, and $n$ if they encircle both. In other words, we recover---up to a numerical factor, and after subtracting the divergence at $\r_2=\r_3$---the two-point function of operators with charge $\hatal_1$, which explains in what sense a {\em continuum} of $\hatal_1$ charges appears in the fusion of charges $\hatal_2$ and $\hatal_3$. 

The loop model also gives natural explanations to a well-known  paradox encountered in trying to make $c \leq 1$ Liouville into a consistent CFT. We see here  that the vertex operator with $\hatal=0$ must be  interpreted not as the identity but as a `marking' operator, a feature very similar to what happens in the SLE construction. For instance, in the CFT proof \cite{CardyHouches} of Schramm's left-passage formula \cite{Schramm} a point-marking (or ``indicator'' \cite{CardyHouches}) operator of zero conformal weight---but distinct from the identity operator---is used to select the correct conformal block in the corresponding correlation function.
Similar arguments can be made within boundary CFT \cite{zigzag}.

We emphasize (more details in SM) that nothing special seems to appear as $c \to 1$. In particular, the loop model does not produce the singularities in the three-point function obtained in \cite{RunkelWatts}. This is, from our point of view, quite expected, since the model in \cite{RunkelWatts} is obtained as the $m\to\infty$ limit of the $A_m$ RSOS models, whose relationship with the loop model is not straightforward, and involves a complicated sum over sectors. 

We hope to report later on the study of $N=4$ point functions, and to explore the geometrical definitions necessary when even more different loop weights are introduced. Boundary extensions of this work also offer tantalizing possibilities.

Similar three-point functions can be defined involving, instead of loops with modified weights, the so-called watermelon or fuseau \cite{WaterM} operators.
In CG parlance, the electric operators $V_{\hatal_i}$ are then replaced by magnetic operators ${\cal O}_{\ell_i}$ inserting $\ell_i$ defect lines.
We have determined numerically the corresponding $\hat{C}$, the simplest of which (all $\ell_i = 2$) determining the probability that three points
lie on the same loop. However, the results do not match formula (\ref{ZamoC}) when the corresponding $\Delta_{\ell_i}$ are entered, and seem to correspond to another continuation of the Liouiville results than the one considered here. This is but one of the many aspects (such as the physical meaning of the pole in the three point function) of this problem that deserve further study. We  hope to report on  all of this soon (some related results have appeared in \cite{YacBen}). 

\subsection*{Acknowledgments}

We thank I.\ Runkel, I.\ Kostov and especially S.\ Ribault for several illuminating discussions. We also thank  G.\ Delfino for a careful reading of the manuscript and explanations of his work \cite{DelfinoViti}.
Support from the Agence Nationale de la Recherche (grant ANR-10-BLAN-0414: DIME) and the Institut Universitaire de France is gratefully acknowledged.

\vfill
\eject

\clearpage
\newpage

\section{Supplementary Material}

\subsection{The function $\Upsilon(x)$}

We first discuss the special function $\Upsilon(x)$ appearing in the expression (\ref{ZamoC}) for the structure constants $\hat{C}$.
Introducing $\hatq\equiv \hatb+{\hatb}^{-1}$ (not to be confused with $\hat{Q}$), it is related to the double gamma function
$\Gamma_2(x|\omega_1,\omega_2)$ via
\begin{equation} \label{Barnes}
 \Upsilon(x) = \frac{1}{\Gamma_{\hatb}(x) \Gamma_{\hatb}(\hatq - x)} \,,
 \quad
 \Gamma_{\hatb}(x) = \Gamma_2(x | \hatb,\hatb^{-1}) \,.
\end{equation}
It obeys the reflection property
\begin{equation}
  \Upsilon(x) = \Upsilon(\hatq-x)
\end{equation}
together with the functional relations
\begin{equation} \label{analyticcont}
  \begin{aligned}
    \Upsilon(\hatq/2) &= 1 \,, \\
    \Upsilon(x+\hatb) &= \gamma(\hatb x) \ \hatb^{1-2\hatb x} \ \Upsilon(x) \,, \\
    \Upsilon(x+\hatb^{-1}) &= 
    \gamma(x\hatb^{-1}) \ \hatb^{-1+2x\hatb^{-1}} \ \Upsilon(x) \,,
  \end{aligned}
\end{equation}
where $\gamma(x)=\Gamma(x)/\Gamma(1-x)$ and $\Gamma$ is the Euler gamma function.

The integral definition (\ref{ups}) of $\Upsilon(x)$ converges only if $|\hatq - 2\ \mathrm{Re}(x)| < \hatq$, i.e., $0 < \mathrm{Re}(x) < \hatq$. Outside this range, the function has to be extended using the foregoing formulae. 

Note that $\Upsilon(x)$ is entire analytic with zeros at
\begin{equation}
 \begin{aligned}
  x &= -m\hatb^{-1}-n\hatb \,, \\
  x &= (m+1)\hatb^{-1}+(n+1)\hatb \,,
  \end{aligned}
\end{equation}
for any $m,n \in \mathbb{Z}^{\ge 0}$.

\subsection{Special cases of $\hat{C}(\hatal_1,\hatal_2,\hatal_3)$}

In the main text we have discussed in details two special cases of our principal result (\ref{ratiocyl}).
In the \underline{first case} (all $n_i$ equal), the structure constant is
\begin{equation}
  C(\hatal,\hatal,\hatal)\propto {\Upsilon^3\left({\hatq\over 2}-{e\hatb^{-1}\over 2}\right)\Upsilon\left({\hatq\over 2}-{3e\hatb^{-1}\over 2}\right)\over
    \Upsilon^{3/2}\left({\hatb}^{-1}(1-e)\right)\Upsilon^{3/2}\left(\hatb^{-1}(1+e)\right)} \,.
  \label{typeI}
\end{equation}
In the \underline{second case} (only one $n_i\neq n$) we have 
\begin{align}
  C(\hatal,0,0)\propto &
  {\Upsilon^2\left({\hatq\over 2}+{e\hatb^{-1}\over 2}\right)\Upsilon\left({3\hatb-{\hatb}^{-1}\over 2}-{e\hatb^{-1}\over 2}\right) }\nonumber\\
  & \times{  \Upsilon\left({3\hatb-{\hatb}^{-1}\over 2}+{e\hatb^{-1}\over 2}\right)\over \Upsilon^{1/2}\left(\hatb+e\hatb^{-1}\right)\Upsilon^{1/2}\left(\hatb-e\hatb^{-1}\right)} \,.
\end{align}

Because of the finite domain of convergence of the integral, analytic continuation is necessary to obtain results for large ranges of the lattice parameters.

Consider for instance the structure constant (\ref{typeI}) for a bulk value $n=1$, as plotted in Fig.~\ref{fig:Caaa}.
As the weight of the special loops $n_i$ is lowered from $n_i=2$, a zero of the second $\Upsilon$ function in the numerator is first encountered for ${\hatq\over 2}-{3e\over 2\hatb}=0$ or $e={1+b^2\over 3}$, corresponding to $n_i=2\cos{1+g\over 3}=2\cos{5\pi\over 9}=-0.34$. Beyond this value, the naive evaluation of the integral gives $\ln\Upsilon=-\infty$, so the naive structure constant is zero. This is overcome by using the second relation in (\ref{analyticcont}) and thus replacing
\begin{eqnarray}
  \Upsilon\left({\hatq\over 2}-{3e\hatb^{-1}\over 2}\right)\to \hatb^{3e{\hatb}^{-2}-1}
  {\Gamma\left({1-\hatb^{-2}\over 2}+{3e\hatb^{-2}\over 2}\right)\over 
    \Gamma\left({1+\hatb^{-2}\over 2}-{3e\hatb^{-2}\over 2}\right)}\nonumber\\
  \times 
  \Upsilon\left({\hatb+3{\hatb}^{-1}\over 2}-{3e\hatb^{-1}\over 2}\right)
\end{eqnarray}

Another remark concerns the regime $n_i > 2$. Although the integral (\ref{ups}) is convergent here, its naive evaluation
using standard software such as {\sc Mathematica} or {\sc Maxima} poses problems, because the integrand decreases fast with $t$. The practical
solution is to constrain the numerical integration to a suitably chosen finite interval, $t \in [0,t_{\rm max}]$, outside which the
integrand is exponentially small.

\subsection{Limit of central charge $c=1$}

The formulae simplify considerably in the limit $c=1$, and allow a comparison with the construction of \cite{RunkelWatts}. The electric charge is then defined by $e=2\hatal$, and the conformal weights of the vertex operator $V_{\hatal}$ read:
\begin{equation}
  \Delta = \bar\Delta = \hatal^2\equiv {e^2\over 4} \,.
\end{equation}
In the regime when $0<e_i<1$ and $e_1+e_2+e_3<2$, the arguments of $\Upsilon$ in~\eqref{ZamoC} all lie in the interval $]0,\hatq=2[$, where the integral representation~\eqref{ups} is valid, and after a change of variable $\beta=e^{-t}$ in~\eqref{ups}, one obtains 
\begin{equation} \label{strct}
  \hat{C}(\hatal_1,\hatal_2,\hatal_3)=\exp[Q(e_1,e_2,e_3)] \,,
\end{equation}
where
\begin{eqnarray}
  Q(x,y,z) &=& \!\! \int_0^1 \!\!\! {{\rm d}\beta\over(-\ln\beta)(1-\beta)^2}\left[2+
  \!\! \sum_{\epsilon=\pm1} (\beta^{\epsilon x}+\beta^{\epsilon y}\right. \nonumber \\
  & & \!\!\! \left.+ \beta^{\epsilon z}) - \!\!\!\! \sum_{\epsilon_x,\epsilon_y,\epsilon_z=\pm 1}
    \!\!\!\! \beta^{(\epsilon_x x+\epsilon_y y+\epsilon_z z)/2}\right] \,. \label{theQfct}
\end{eqnarray}
%
The structure constants for other values of the electric charges can be related to the above regime through the functional relations~\eqref{analyticcont}.
Note also that for $\hatb=1$, the double gamma function in~\eqref{Barnes} is related to the Barnes $G$-function by
\begin{equation}
  \Gamma_2(x|1,1) = \mathrm{const} \times \frac{(2\pi)^{x/2}}{G(x)} \,.
\end{equation}

The case $c=1$ is particularly interesting since the  loop model then maps, in the standard Coulomb gas formalism \cite{Nienhuis,JesperReview}, to a free boson with no electric charge at infinity. This is because loops in the bulk get a weight $n=2$, which can be obtained simply by summing over two possible orientations. The standard phenomenology \cite{FodaNienhuis} then does not suggest the presence of ``floating charges''. Nonetheless, the Liouville description is again fully confirmed by the numerical study of the loop three-point functions.

\begin{figure}
  \vskip-0.6cm
  \includegraphics[scale=.34]{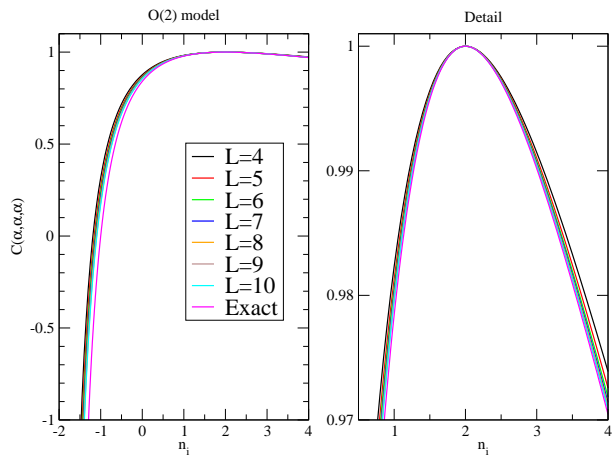}
  \vskip-0.9cm
  \caption{$\hat{C}(\hatal,\hatal,\hatal)$ as a function of $n_1=n_2=n_3$ in the dense O($n$) model with $n=2$, compared with the $c=1$ limit of~\eqref{ZamoC}.}
  \label{fig:Caaa2}
\end{figure}

We illustrate this by considering  the \underline{first case}  of three-point couplings for identical loop fugacities, $\hat{C}(\hatal,\hatal,\hatal)$.
As seen in~\figref{Caaa2}, at the point $n_1=n_2=n_3=-1$, corresponding to $e_1=e_2=e_3=2/3$ where \eqref{theQfct} diverges, the structure constant observed numerically in the loop model behaves smoothly, as predicted by the Liouville description. This contrasts with the results of~\cite{RunkelWatts}, where the unitarity constraints in minimal models are reflected as a non-analytic behaviour of $\hat C$ in the limit $c \to 1$, which would produce $\hat C=0$ all along the range $-2<n_1=n_2=n_3<-1$. This particular example shows that the $c \to 1$ limit of the Liouville action~\eqref{nuL} and that of minimal models lead to different theories, and that the loop model clearly relates to the Liouville theory.

\subsection{Orthogonality catastrophe}

A potential source of confusion must be dispelled at this stage. Indeed, the way we determine numerically the three-point function on the cylinder is reminiscent of the calculation of overlaps of ground states for twisted spin chains. While one has to be careful with this correspondence in general, it becomes exact in the case $n=2$ and  $\hatal_2=\hatal_1+\hatal_3$, where the 
partition function used to determine $\hat{C}$ is proportional to the overlap of XXX ground states with two different twists. When $\hatal_1 \neq \hatal_3$, this overlap vanishes when the radius of the cylinder $L\to\infty$ as a power law of $L$, a manifestation of the well-known Anderson orthogonality catastrophe \cite{Anderson67}. However, this power law is factored out by the normalization (see below), which extracts directly the corresponding amplitude. Of course, this amplitude is trivial: one can check that $\hat{C}(\hatal_1,\hatal_1+\hatal_3,\hatal_3)=1$, since $Q(x,x+z,z)=0$, as follows from the simple identity ($l\equiv -\ln\beta$)
\begin{eqnarray}
8\cosh {lx\over 2}\cosh{ ly\over 2}\cosh {l(x+y)\over 2}=\nonumber\\
2\left[1+\cosh lx+\cosh ly+\cosh l(x+y)\right] \,.
\end{eqnarray}

We further emphasize that the $\hat{C}$ function obtained in the loop model does not exhibit any of the singularities observed in the limit of minimal models studied in  \cite{RunkelWatts}: everything behaves as if the prefactor $P(x,y,z)$ in this reference were simply absent. 

\subsection{Loop models on the square lattice}

The loop model on the square lattice has nine possible configurations
at each vertex:

\begin{center}
  \begin{tikzpicture}[scale=0.55]
    \begin{scope}[rotate=45]
      \draw [black, dashed,line width=0.2] (-0.5,-0.5) -- (0.5,0.5); 
      \draw [black, dashed,line width=0.2] (0.5,-0.5) -- (-0.5,0.5); 
    \end{scope}
    \node at (0,-1.) {$\rho_1$};
    \begin{scope}[shift={(1.75,0)}]
      \begin{scope}[rotate=45]
        \draw [black, dashed,line width=0.2] (-0.5,-0.5) -- (0.5,0.5); 
        \draw [black, dashed,line width=0.2] (0.5,-0.5) -- (-0.5,0.5); 
        \draw[purple,line width=2,rounded corners=5pt]%
        (-0.5,-0.5) -- (0,0) -- (-0.5,0.5);
      \end{scope}
      \node at (0,-1.) {$\rho_2$};
    \end{scope}
    \begin{scope}[shift={(3.5,0)}]
      \begin{scope}[rotate=45]
        \draw [black, dashed,line width=0.2] (-0.5,-0.5) -- (0.5,0.5); 
        \draw [black, dashed,line width=0.2] (0.5,-0.5) -- (-0.5,0.5); 
        \draw[purple,line width=2,rounded corners=5pt]%
        (0.5,-0.5) -- (0,0) -- (0.5,0.5);
      \end{scope}
      \node at (0,-1.) {$\rho_3$};
    \end{scope}
    \begin{scope}[shift={(5.25,0)}]
      \begin{scope}[rotate=45]
        \draw [black, dashed,line width=0.2] (-0.5,-0.5) -- (0.5,0.5); 
        \draw [black, dashed,line width=0.2] (0.5,-0.5) -- (-0.5,0.5); 
        \draw[purple,line width=2,rounded corners=5pt]%
        (-0.5,-0.5) -- (0,0) -- (0.5,-0.5);
      \end{scope}
      \node at (0,-1.) {$\rho_4$};
    \end{scope}
    \begin{scope}[shift={(7,0)}]
      \begin{scope}[rotate=45]
        \draw [black, dashed,line width=0.2] (-0.5,-0.5) -- (0.5,0.5); 
        \draw [black, dashed,line width=0.2] (0.5,-0.5) -- (-0.5,0.5); 
        \draw[purple,line width=2,rounded corners=5pt]%
        (-0.5,0.5) -- (0,0) -- (0.5,0.5);
      \end{scope}
      \node at (0,-1.) {$\rho_5$};
    \end{scope}
    \begin{scope}[shift={(8.75,0)}]
      \begin{scope}[rotate=45]
        \draw [black, dashed,line width=0.2] (-0.5,-0.5) -- (0.5,0.5); 
        \draw [black, dashed,line width=0.2] (0.5,-0.5) -- (-0.5,0.5); 
        \draw[purple,line width=2,rounded corners=5pt]%
        (-0.5,-0.5) -- (0,0) -- (0.5,0.5);
      \end{scope}
      \node at (0,-1.) {$\rho_6$};
    \end{scope}
    \begin{scope}[shift={(10.5,0)}]
      \begin{scope}[rotate=45]
        \draw [black, dashed,line width=0.2] (-0.5,-0.5) -- (0.5,0.5); 
        \draw [black, dashed,line width=0.2] (0.5,-0.5) -- (-0.5,0.5); 
        \draw[purple,line width=2,rounded corners=5pt]%
        (0.5,-0.5) -- (0,0) -- (-0.5,0.5);
      \end{scope}
      \node at (0,-1.) {$\rho_7$};
    \end{scope}
    \begin{scope}[shift={(12.25,0)}]
      \begin{scope}[rotate=45]
        \draw [black, dashed,line width=0.2] (-0.5,-0.5) -- (0.5,0.5); 
        \draw [black, dashed,line width=0.2] (0.5,-0.5) -- (-0.5,0.5); 
        \draw[purple,line width=2,rounded corners=5pt]%
        (-0.5,-0.5) -- (0,0) -- (-0.5,0.5);
        \draw[purple,line width=2,rounded corners=5pt]%
        (0.5,-0.5) -- (0,0) -- (0.5,0.5);
      \end{scope}  
      \node at (0,-1.) {$\rho_8$};
    \end{scope}
    \begin{scope}[shift={(14,0)}]
      \begin{scope}[rotate=45]
        \draw [black, dashed,line width=0.2] (-0.5,-0.5) -- (0.5,0.5); 
        \draw [black, dashed,line width=0.2] (0.5,-0.5) -- (-0.5,0.5); 
        \draw[purple,line width=2,rounded corners=5pt]%
        (-0.5,-0.5) -- (0,0) -- (0.5,-0.5);
        \draw[purple,line width=2,rounded corners=5pt]%
        (-0.5,0.5) -- (0,0) -- (0.5,0.5);
      \end{scope}
      \node at (0,-1.) {$\rho_9$};
    \end{scope}
  \end{tikzpicture}
\end{center}

Each vertex has a Boltzmann weight $\rho_i$ as indicated, and there is a weight $n$ for
each closed loop. In the computation of three-point functions, some of the loops will get
modified weights $n_1$, $n_2$ or $n_3$ (see below).

The square-lattice O($n$) model referred to in the main text corresponds to a set of
integrable weights \cite{Nienhuis89} which read, at the isotropic point,
\begin{eqnarray}
 \rho_1 &=& 1 + \sin \lambda + \sin(3 \lambda) - \sin(5 \lambda) \,, \nonumber \\
 \rho_2 &=& \rho_3 = \rho_4 = \rho_5 = 2 \sin(2 \lambda) \sin((6 \lambda + \pi)/4) \,, \nonumber \\
 \rho_6 &=& \rho_7 = 1 + \sin(3 \lambda) \,, \\
 \rho_8 &=& \rho_9 = \sin \lambda + \cos(2 \lambda) \,, \nonumber \\
 n &=& -2 \cos(4 \lambda) \,. \nonumber
\end{eqnarray}
The dense (resp.\ dilute) phase is obtained for the parameter range $0 < \lambda \le \frac{\pi}{4}$
(resp.\ $\frac{\pi}{4} \le \lambda \le \frac{\pi}{2}$), and the corresponding CG coupling constant is
$g = \frac{4 \lambda}{\pi} \in (0,2]$.

The loop model which is equivalent to the $Q$-state Potts model \cite{BKW76}
corresponds to another choice of integrable weights:
\begin{eqnarray}
 \rho_1 &=& \cdots = \rho_7 = 0 \,, \nonumber \\
 \rho_8 &=& \rho_9 = 1 \,, \\
 n &=& \sqrt{Q} = -2 \cos(\pi g) \,, \nonumber
\end{eqnarray}
so that each edge is covered by a loop (complete packing).
The CG coupling $g \in (0,1]$ covers only the dense phase in this case.

\subsection{Transfer matrix method for measuring $\hat{C}(\hatal_1,\hatal_2,\hatal_3)$}

\subsubsection{General setup}

Let $(\Phi_1, \Phi_2, \Phi_3)$ be three quasi-primary operators of a CFT, for which we want to compute the structure constant $\hat{C}(\Phi_1, \Phi_2, \Phi_3)$. The idea of our method is to consider the problem on the cylinder, and use the operator/state correspondence with operators and states normalized so that $\Phi_j\ket{0} = \ket{\Phi_j}$ (where $\ket 0$ is the ground state) to write
\begin{equation}
  \hat{C}(\Phi_1, \Phi_2, \Phi_3) = \aver{\Phi_1|\Phi_2|\Phi_3} \,.
\end{equation}
In order to achieve a finite-size estimation of this expression, we simply need two ingredients: (i) the eigenstates $\ket{\Phi_1}_L$ and $\ket{\Phi_3}_L$ of the transfer matrix for the lattice model on the cylinder of circumference $L$, which converge to $\ket{\Phi_1}$ and $\ket{\Phi_3}$ as $L \to \infty$; and (ii) the average value of the finite-size operator $\Phi_2^{(L)}$ between two basis states $\bra a$ and $\ket b$. The latter operator will typically scale like $\Phi_2^{(L)} \sim \mathcal{N}_2 L^{-\Delta_2-\bar\Delta_2} \ \Phi_2$, where $\mathcal{N}_2$ is a non-universal factor. This undetermined normalization is then easily eliminated by taking the ratio:
\begin{equation} \label{eq:tm}
  \frac{_L\aver{\Phi_1|\Phi_2^{(L)}|\Phi_3}_L}
  {_L\aver{\Phi_2|\Phi_2^{(L)}|0}_L} \ \mathop{\longrightarrow}_{L \to \infty} \ \hat{C}(\Phi_1, \Phi_2, \Phi_3) \,.
\end{equation}
The above relation gives a direct way to access the constant $\hat{C}(\Phi_1, \Phi_2, \Phi_3)$, by evaluating the eigenstates $\ket{\Phi_1}_L$, $\ket{\Phi_3}_L$ and $\ket{0}_L$, acting with $\Phi_2^{(L)}$, and performing the required scalar products. The convergence in $L$ is typically algebraic.

\subsubsection{Time-efficient version}

In loop models, the most time-consuming part in this calculation is computing the scalar product, which takes O($D_L^2$) operations, where $D_L$ is the dimension of the Hilbert space for the transfer matrix, growing exponentially in $L$. This is because the scalar product $\aver{a|b}$ between two basis states is always non-zero. In the case when both $\ket{\Phi_1}$ and $\ket{\Phi_3}$ are the lowest-energy states in two symmetry sectors of the transfer matrix, the method can be improved as follows. Denoting $t_j$ the transfer matrix in the sector where $\ket{\Phi_j}_L$ lives, and $\ket{a_j}_L$ a basis state in this sector with $_L\aver{a_j|\Phi_j}_L \neq 0$, we use the fact that
\begin{equation*}
  \frac{_L\aver{a_1|t_1^M \Phi_2^{(L)} t_3^M |a_3}_L}
  {\sqrt{_L\aver{a_1|t_1^{2M}|a_1}_L \ _L\aver{a_3|t_3^{2M}|a_3}_L}} \ \mathop{\longrightarrow}_{M \to \infty} \ _L\aver{\Phi_1|\Phi_2^{(L)}|\Phi_3}_L
\end{equation*}
for {\em finite} cylinder of $2M$ rows, where $M \gg L$.
In practice, due to the exponential convergence with respect to $M$, only $M \sim 10L$ iterations are needed to reach the limit within machine precision. This allows us to evaluate each scalar product on the left-hand side of~\eqref{eq:tm} in just O($D_L$) operations, while using essentially the same amount of memory as in the direct method.

\subsubsection{Case of three purely electric operators}

\begin{figure}
  \begin{center}
    \includegraphics{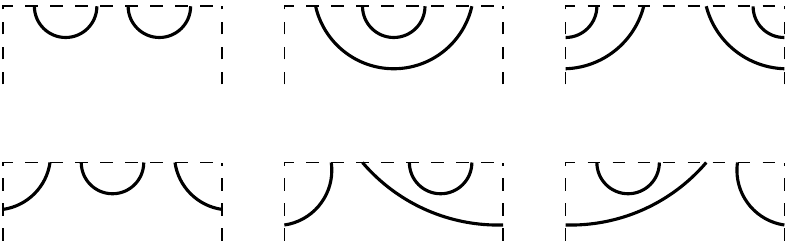} \\
    (a) \\ \bigskip
    \includegraphics{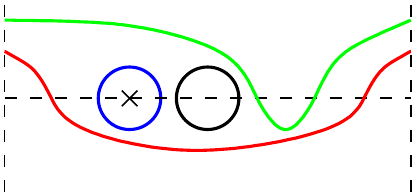} \\
    (b)
  \end{center}
  \caption{(a) Link patterns for a transfer matrix of size $L=4$. 
  We illstrate the Potts case (complete packing); the O($n$) model further admits empty points (dilution). 
  (b) Gluing link patterns $a$ and $b$ is defined by juxtaposing $a$ and the reflection of $b$ in a horizontal mirror. The loop weights are indicated with the same colors as in \figref{loopconfs}.}
  \label{fig:states}
\end{figure}

Let us now explain how the above method is applied to the numerical computation of $\hat{C}(\hatal_1,\hatal_2,\hatal_3)$. Our transfer matrix acts on the space of non-crossing link patterns between $L$ points, where some arcs can pass through the periodic boundary condition horizontally (see \figref{states}a). The sector of the transfer matrix defined by giving a weight $n_i$ to non-contractible loops contains (the discrete analog of) the Verma modules of $\ket{V_{\hatal_i}}$ with 
\begin{equation} \label{eq:charge}
  \hatal_i \in \frac{\hatQ}{2} + \frac{e_i + 2\mathbb{Z}}{2 \hatb} \,,
\end{equation}
where $e_i = \mathrm{Arccos}(n_i/2) / \pi $. So we pick a state $\ket{V_{\hatal_i}}_L$ in the sector of weight $n_i$, for $i=1,2,3$. The operator $\Phi_2^{(L)}(j)$, associated to loop weight $n_2$, and sitting at site $j \in \{1, \dots, L\}$, is defined through its average value between two basis states:
\begin{equation}
  \aver{a_1|\Phi_2^{(L)}(j)|b_3} = n^{\ell(a,b)} n_1^{\ell_1(a,b)} n_2^{\ell_2(a,b)} n_3^{\ell_3(a,b)} \,,
\end{equation}
where, after gluing the states $a$ and $b$ (see \figref{states}b), $\ell(a,b)$ (resp. $\ell_2(a,b)$) is the number of contractible loops not surrounding $j$ (resp. surrounding $j$), and $\ell_1(a,b)$ (resp. $\ell_3(a,b)$) is the number of non-contractible loops below $j$ (resp. above $j$). The lattice operator $\Phi_2^{(L)}(j)$ contains contributions from the $V_{\hatal_2}$'s with $\hatal_2$ defined by \eqref{eq:charge} with $i=2$, and their descendants under the Virasoro algebra:
\begin{equation}
  \Phi_2^{(L)} \sim \sum_{e'_2 \in e_2+2\mathbb{Z}} \mathcal{N}(e'_2) \ L^{-2\Delta(e'_2)} \ V_{\hatQ/2 + e'_2/(2\hatb)} + \mathrm{desc.}
\end{equation}
where $\mathcal{N}(e'_2)$ is an undetermined prefactor. Thus, as in \eqref{eq:tm}, the ratio
$$
\frac{_L\aver{V_{\hatal_1}|\Phi_2^{(L)}|V_{\hatal_3}}_L}{_L\aver{V_{\hatal_2}|\Phi_2^{(L)}|0}_L}
$$
converges to $\hat{C}(\hatal_1,\hatal_2,\hatal_3)$ for $e_2 \in [-1,1]$. The value of $\hat{C}$ for a charge $e'_2=e_2+2k$ with $k \in \mathbb{Z}$ can be obtained by picking the correct state $\ket{V_{\hatal_2}}_L$ and selecting the subdominant term with a prescribed scaling $L^{-2\Delta(e_2+2k)}$ in the numerator and denominator of~\eqref{eq:tm}.

\subsubsection{Algorithm without scalar products}

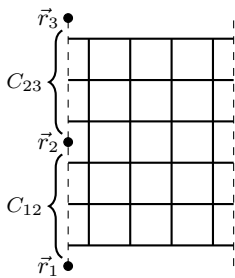
\begin{figure}
 \begin{center}
 \begin{tikzpicture}[scale=0.55]
  \foreach \ypos in {0,1,2,3,4,5}
    \draw[thick] (0,\ypos)--(4,\ypos);
  \foreach \xpos in {0.5,1.5,2.5,3.5}
   \draw[thick] (\xpos,0)--(\xpos,5);
 \draw[dashed] (0,-0.5)--(0,5.5);
 \draw[dashed] (4,-0.5)--(4,5.5);
 \filldraw (0,-0.5) circle(3pt);
 \filldraw (0,2.5) circle(3pt);
 \filldraw (0,5.5) circle(3pt);
 \draw (0,-0.5) node[left]{$\r_1$};
 \draw (0,2.5) node[left]{$\r_2$};
 \draw (0,5.5) node[left]{$\r_3$};
 \draw [thick,decorate,decoration={brace,amplitude=5pt},xshift=-4pt,yshift=0pt] (0,-0.3) -- (0,2.2)
  node [black,midway,xshift=-0.5cm] {\footnotesize $C_{12}$};
 \draw [thick,decorate,decoration={brace,amplitude=5pt},xshift=-4pt,yshift=0pt] (0,2.7) -- (0,5.2)
  node [black,midway,xshift=-0.5cm] {\footnotesize $C_{23}$};
 \end{tikzpicture}
 \end{center}
 \caption{Cylinder geometry with $L=4$ and $M=3$. The dashed lines are identified by the periodic boundary condition.}
 \label{fig:sqlatt}
\end{figure}

The time-efficient version of our algorithm thus computes the partition functions
$Z_{ijk} \equiv Z_{n_i,n_j,n_k}(\r_1,\r_2,\r_3)$ on a finite cylinder of the square lattice,
with circumference $L$ and height $2M$ rows, in the limit $M \gg L$. The three marked
points reside on lattice faces, with $\r_1$ (resp.\ $\r_3$) situated just below (resp.\ above)
the bottom row (resp.\ top) row, and $\r_2$ in the middle (see Fig.~\ref{fig:sqlatt}).
For the Potts model we must take both $L$ and $M$ even to respect the sublattice
parities in the mappings from the spin model \cite{BKW76}, whereas for the O($n$)
model there are no such parity effects.

The states of the transfer matrix are of the type shown in Fig.~\ref{fig:states}a, but with the possibility of dilution (empty sites) in the O($n$) model. The boundary condition below the bottom row is taken as the first state in Fig.~\ref{fig:states}a; in the Potts case this corresponds to free boundary conditions for the Potts spins. A similar boundary condition (horizontally reflected) is applied above the top row. Note that with these boundary conditions neither $\r_1$ nor $\r_3$ can be be surrounded by a loop. The structure constant $\hat{C}(\hatal_1,\hatal_2,\hatal_3)$ is then computed from $Z_{ijk}$ as explained around (\ref{ratiocyl}) in the main text.

To build the lattice efficiently by the transfer matrix, we factorize the latter as a product of sparse matrices $\check{R}$, each corresponding to the addition of one vertex. The states are organized in a hash table. Each arc in the states (see Fig.~\ref{fig:states}a) carries two binary variables $N_{12} = 0,1$ (resp.\ $N_{23}$) that count the number of times {\em modulo 2} that the arc has intersected the cut $C_{12}$ (resp.\ $C_{23}$) running from $\r_1$ to $\r_2$ (resp.\ from $\r_2$ to $\r_3$). The first stage in building a row with the transfer matrix is the insertion of the auxiliary space, i.e., the first horizontal edge that crosses $C_{12}$ (resp.\ $C_{23}$) on the lower (resp.\ upper) half cylinder. If that edge carries an arc, we initialize the corresponding $N_{12} = 1$ (resp.\ $N_{23} = 1$). In all other stages of the transfer process, when two arcs are joined their variables $N_{12}$ and $N_{23}$ add up {\em modulo 2}. Finally, a loop resulting from the closure of an arc is attributed the weight $n_i$ given by (\ref{loop weights}) in the main text (cf.\ Fig.~\ref{fig:states}b).

\subsubsection{Finite-size extrapolations}

In Figs.~\ref{fig:Caaa}, \ref{fig:Ca00} and \ref{fig:Caaa2} we have displayed only the raw results (\ref{ratiocyl}) for sizes $L=4,5,\ldots,10$. The convergence towards the analytical result---which is already convincing by visual inspection---can be further improved by computing finite-size scaling (FSS) extrapolations
\begin{equation}
  \hat{C}(L) = \hat{C}(\infty) + \gamma L^{-k}
\end{equation}
from three consecutive sizes $L$. For instance, in the case of Fig.~\ref{fig:Caaa} the FSS exponent $k$ was found to increase monotonically with $n_i$---from 1.1 to 2.0 (resp.\ 0.6 to 2.1) in the dense (resp.\ dilute) case, for the parameter values shown---except for $n_i = 1$ where $\hat{C} = 1$ exactly for any $L$. The resulting agreement between $\hat{C}(\infty)$ and (\ref{ZamoC}) amounts to at least four significant digits over the whole parameter range.

\subsection{The factor $\sqrt{2}$ in the Potts model}

The correspondence between the lattice discretization and the continuum limit is somewhat singular in the case when non-contractible loops on the cylinder get zero weight (i.e., $n_1 = 0$ and/or $n_3 = 0$). The states on which the transfer matrix acts are of two types, or parities, depending on whether the number of arcs crossing the periodic boundary condition is even or odd. For instance, in Fig.~\ref{fig:states}a the three states in the first (resp.\ second) line are even (resp.\ odd). The two subsets are related by a shift by one lattice spacing, and thus have the same cardinality; this relationship corresponds to the Kramers-Wannier duality of the Potts model.

It is easy to see \cite{J14,GRS} that when non-contractible loops are forbidden the parity just defined is conserved by the transfer matrix. In more mathematical terms, the transfer matrix acts on a representation of the periodic Temperley-Lieb algebra which is generically irreducible, but breaks down into two irreducible, isomorphic sub-representations when $n_i=0$. The lattice procedure we use restricts to just one of these sub-representations, while by continuation of the general $n_i$ case, the theoretical formula for the three-point function should involve, in the lattice regularization, the sum of both sub-representations.  Associate formally to each of these representations two fields with the same conformal weight (but vanishing cross two-point function) $V_1$ and $V_2$. The theoretical formula, considering for instance the case where the three $n_i=0$, applies formally to $V\equiv {V_1+V_2\over \sqrt{2}}$ (the $\sqrt{2}$ factor arises from the normalization of the two-point function). In obvious short-hand notations one has $VVV={V_1V_1V_1\over\sqrt{2}}$. If in the transfer matrix calculation we measure $V_1V_1V_1$, we see that the result should be $\sqrt{2}$ larger than the theoretical value (which corresponds to $VVV$), in agreement with the numerical observations.

\end{document}